\documentstyle[12pt,epsf]{article}
\font\tenrm=cmr10
\font\tenit=cmti10

\begin{document}

\font\fortssbx=cmssbx10 scaled \magstep2
\hbox to \hsize{

\hfill$\vcenter{\hbox{\tt hep-ph/9709269 } }$ }

\vspace{.1in}

\begin{center}
{\LARGE \bf A New Signature of Dark Matter}
\\
\vskip 0.7cm
{\bf {\tenrm S.~Abbas }} \\
{\tenit Physics Dept., Utkal University, Bhubaneswar-751004, India \\
e-mail: abbas@iopb.res.in } 
\footnote{ Address
for correspondence : Institute of Physics,
Bhubaneswar-751005, Orissa, India} 
\\[.1cm]
{\bf {\tenrm A.~Abbas}} \\
{\tenit Institute of Physics,
 Bhubaneswar-751005, Orissa, India \\
 e-mail : afsar@iopb.res.in } 
\\[.1cm]
{\bf {\tenrm S.~Mohanty}} \\
{\tenit Physics Dept., Utkal University, Bhubaneswar-751004, India}
\end{center}

\vskip 4 mm
\centerline{\bf{PACS Numbers}: 95.35.+d, 98.80.Cq }
\vskip 0.7 cm

\smallskip


\begin{center}ABSTRACT \end{center}


Dark matter capture and annihilation in planets and their 
satellites, in addition to producing neutrinos (already the basis 
of several ongoing experiments), is shown to lead to more significant
heat generation in these bodies for a uniform dark matter halo. 
This thermal output becomes more prominent when clumped dark matter 
passes through the solar system. The dark matter annihilations should be
treated as a new source of heat in the 
solar system which in some special cases may lead to unique imprints. Such
new signatures of 
the dark matter are found in the generation of the puzzling 
magnetic field of Ganymede. This new source of heat perhaps cannot explain
the
formation and segregation of Ganymedean structure (which are
probably due to conventional gravitational and radioactive heating), but
provides a possible explanation of the origin of the Ganymedean magnetic
field.


\newpage

\section{ Introduction }

Investigations have revealed that clumping \cite{silk} may 
be a generic characteristic of dark matter \cite{kolb, hogan}. 
The weakly interacting massive particles ( WIMPs ), an 
especially favoured group of dark matter candidates \cite{jung} 
that are thought to comprise the galactic halo may be
detected by direct methods or indirect techniques \cite{jung}.
Nuclear recoil from elastic WIMP - nucleus collisions is used
in the direct technique. The indirect detection procedures utilize
solar \cite{press} and terrestrial \cite{krauss, gould} 
capture of these WIMPs. After crossing a critical threshold 
density \cite{freese} these particles start annihilating with one 
another \cite{gaisser}. The neutrinos escaping from the centre of 
the Earth and Sun through this process have been the subject of 
intensive investigation \cite{jung, bengston, bere, rich}. 
So far the searches as reported by Kamiokande, IMB, Frejus 
and MACRO have yielded null results \cite{jung}. In this paper 
it shall be pointed out that in addition to neutrinos the above 
dark matter annihilations produce large amounts of heat as well.
This effect should be treated as a new source of heat in the solar
system. In some special situations where the conventional heat
sources fail to explain a particular phenomenon dark matter
heat generation may leave a singular imprint. 
We shall discuss one such possible signature from recent puzzling 
and unexpected observations in the solar system.

 The conventional accretional and radiogenic heat sources are known to be
important during the early phases of a planet's history. Thus, whenever
new empirical information arises indicating more recent heat sources (eg.
the presence of Ganymede's magnetic field), the tendency of planetary
scientists is to seek refuge in the only other known source of heat, ie.
tidal heating. Even when the orbital parameters constrain the present
tidal heating to insignificant proportions, sometimes contrived and
artificial models are propagated to explain mysterious emirical
information. In this paper it is asserted that, in addition to the
conventional sources of heat mentioned above, dark matter annihilation in
planetary interiors may be a new heat source, in addition to explaining
several mysteries in planetary science.

\section{ Requirement of New Sources of Heat }

The first such mystery is the absolutely unexpected discovery
during the recent Galileo mission of the magnetic field of 
Ganymede, the largest satellite of Jupiter \cite{kiv} which 
necessitates a reevaluation of the structure of Ganymede 
\cite{and}. This mission uncovered the existence of a surface
equatorial magnetic field of  $ \sim 750 nT $ , strong enough to carve 
out a magnetosphere with clearly defined boundaries within Jupiter's.
Since magnetic fields generally arise via a 
dynamo action \cite{schu} this observation implies a partly
molten iron or iron-iron sulphide core. 
Accretional heating is probably inadequate for early 
formation of the Ganymedean metallic core, though it may be 
sufficient to segregate the ice from the rock and metal \cite{schu}.
Radiogenic heating during the course of Ganymede's 
evolution however could provide enough heat to separate an Fe-FeS core 
if the eutectic melting temperature of 1250K is all that is 
required. Thus formation of a metallic core in Ganymede 
may not be energetically difficult, but there is 
a more serious problem involved in explaining how the core has 
remained convective throughout Ganymede's history 
as small cores tend 
to evolve toward an isothermal state \cite{schu}. 
Speculative solutions within the 
tidal heating framework involve a jolt of tidal heating in
the past, but these are at best contrived and rather artificial 
\cite{kerr}. It may be reasonable to conclude that  
a consistent physical explanation of the origin of the
Ganymedean magnetic field does not exist. 
This may be an indication of `new 
physics'. It will be shown below that this new physics may be
provided by dark mattter.

\section{ Critical Threshold and Capture Rates }

In the cosmic string, texture and inflationary models
dark matter in a clumped form is a natural outcome. 
The concentration of dark matter inside these clumps can 
reach $ 10^{ 10 } - 10^{14} $ times the background for
WIMPs \cite{silk, kolb}. The dark matter particles continue 
to accumulate inside the planet until their number density 
becomes so high that they start to annihililate with one 
another. Let this critical number be $ N_{\oslash}^{crit} $. 
Equilibrium is achieved when the capture rate equals the 
annihilation rate. The critical number density of sneutrinos 
for any planet is \cite{freese}

\begin{equation}
N_{\oslash}^{crit} = \sqrt{ \frac{ 4\pi }{3} } 
      \{ r_{ \tilde{\nu} }^{3/2} \}
      \left[ \sqrt{ \dot{ N_{\oslash } } } \right]
	\left( \frac{1}{ (\sigma \, v)_{ann} } \right)^{1/2}
\end{equation}

where $\dot{ N_{\oslash} }$ is the 
capture rate for the planet in question, $ r_{ \tilde{\nu} } $
is the scale radius of sneutrinos inside the core, 
$v$ the dispersion velocity of the dark matter and 
$(\sigma \, v )_{ann}$ the 
annihilation cross-section.
The critical number for Ganymede in the case of
an iron core is therefore

\begin{equation}
N_{G}^{crit} = 9.1 \times 10^{33} 
    \sqrt{\Omega_{\tilde{\nu}} } \, h_{50}  
    \left( \frac{ m_{ \tilde{\nu} } }{ 10 GeV } \right)^{-3/4}
    \sqrt{ \frac{ \dot{ N_{G} } }{ 10^{17} sec^{-1} } }
\end{equation}

which compares to a critical value for the Earth of \cite{freese}

\begin{equation}
N_{\oplus}^{crit} = 1.7 \times 10^{34} 
    \sqrt{\Omega_{\tilde{\nu}} } \, h_{50}      
    \left( \frac{ m_{ \tilde{\nu} } }{ 10 GeV } \right)^{-3/4}
    \sqrt{ \frac{ \dot{ N_{\oplus} } }{ 10^{17} sec^{-1} } }
\end{equation}

where $ h_{50} $ is the normalized Hubble constant and 
$\Omega_{ \tilde{\nu} } $ is the ratio of the 
sneutrino density to the closure density. 
These and the following formulas hold for sneutrinos 
but similar results hold for most WIMPs.


The time period between successive core passings is 
$ \tau = \left( y \cal{F} \sigma \right)^{-1}  $ 
where $\cal{F}$ is the flux of the DM particles, $\sigma$ 
the cross-section and $y$ the fraction of dark matter still 
in clump cores. Hence 
$ \tau = y^{-1} \left( n \sigma v \right)^{-1} $ i.e.,
$ \tau = M_{C} ( y \rho_{H} v {\sigma} )^{-1} $.
Since the core mass \\
$ M_C \sim \rho_{C} R_{C}^3 $, 
and $ \sigma \sim R_{C}^2 $, one obtains
$ \tau = (\rho_{C} R_C)(\rho_{H} v)^{-1} $. \\
Since $ M_C = 0.02 M_{\odot} 
          \left( m_{X}^{3/2} \Omega_0 h^2 \right)^{-1} $, 
$ \sigma = \pi R^2 $ with
$ R = m_{X}^{-1/2} \Omega_{0}^{-1} h^{-2} \times 10^{-3} pc $
and taking $v$ to be 300 km/sec the relation between 
the mass of the dark matter and the time period of crossing is

\begin{equation}
\tau = 2.1 \times 10^9 m_{X}^{-1/2} y^{-1} 
       \Omega_0 h^2 years
\end{equation}

where $m_{X}$ is the mass number of the dark matter 
particle in GeV.

We thus find \cite{collar}
that a time scale of $ \sim 10 - 100 My $ results. 
Hence, as explained below, Ganymede may have experienced a 
jolt of rejuvenation approximately $ 10 - 100 My $ ago, perhaps 
leading to the melting of a 
part of Ganymede's core with the consequent rebirth 
or reinvigoration of a magnetic field. 


Improving upon the previous work, Gould 
\cite{gould} 
obtained greatly enhanced capture rates for the Earth 
( 10-300 times that previously believed ) when the WIMP mass 
roughly equals the nuclear mass of an element present in the 
Earth in large quantities, thereby constituting a resonant 
enhancement. Gould's formula gives the capture rate for each 
element in any planet as \cite{gould} :



\begin{equation}
\dot{N} =  \left[  \sqrt{ \frac{8}{3 \pi} } \sigma n_x \bar{v} \right]
           \frac{ M }{ m_N }
           \left[ \frac{ 3 v_{esc}^2 }{ 2 \bar{v}^2 }  \right]
           < \hat{\phi} >
               \left[ \xi_{\eta}( \infty ) \right] 
                  \left< \frac{ \hat{\phi} }{ <\hat{\phi}> }
                         \left( 1 - \frac{ 1-e^{-A^2} }{ A^2 } \right) 
                         \frac{ \xi_{\eta}(A) }{ \xi_{\eta}(\infty) }
                          \right> 
\end{equation}

 where the third term is the escape velocity term. $ v_{esc} $ is the
escape velocity, Dirac brackets indicate averaging over the mass of the
body, M is the mass of the body and
$ \xi_{1} (A) $ is a correction factor, and $ \hat{\phi} $ is the
dimensionless gravitational potential.


To apply the Gould formula
to Ganymede, we note that 

\begin{itemize}
  \item The mass of Ganymede as determined by the Galileo mission is  $
1.482 \times 10^{23} kg $ \cite{and}.
  \item $ v_{esc} = 2.75 km/sec $, $ \bar{v} = 300 km/sec $
  \item $ n_x = ( 1/m_x )  \rho_{0.4} 0.4~Gev/cm^2 $
  \item $ \sigma = \frac{\mu}{ \mu_{+}^2 }  Q^2  
                   \frac{ m_X m_N }{ (GeV)^2 }
          ( 5.2 \times 10^{-40} cm^{-2} ) $
\end{itemize}

This yields the Ganymedean capture rate as :

\begin{equation}
 \dot{N_G} = ( 6 \times 10^{13} sec^{-1} ) 
             \left[ \frac{\mu}{ \mu_{+}^2 } Q^2 f \rho_{0.4} \right]
            \left< \hat{\phi} (1 - \frac{1 - e^{-A^2} }{ A^2 } )
                   \xi_{1}(A) \right>
\end{equation}

Then, noting that

\begin{itemize}
  \item $ Q = N - ( 1 - 4  sin^2 \theta_W  ) Z = 30 - 0.124 \times 26 =
    26.8$ ,
  \item The mass of Ganymede's core ranges from 
$ 2\%-33\%$ of the planetary mass in case of Fe-FeS 
core and from $ 1.4\%-26\%$ in the case
of an Fe core \cite{and}.
  \item For $ A >> 1 $ , the last term in the Dirac brackets is one.

\end{itemize}

This yields a capture rate of 

\begin{equation}
\dot{ N_{G} }^{Fe} =  4.33 f \times 10^{16} s^{-1} 
\end{equation}

where $f$ = fraction of Fe in the Fe-FeS core and varies from $ 2 \% $
to $ 33 \% $ for an Fe-FeS core.


One obtains capture rates of $ \sim 10^{14} - 10^{16} s^{-1}$ particles
per sec,
depending on the quantity of iron.
For dark matter masses other than 52 GeV, 
but lying in the range of 10-100 GeV,
the capture rates are generally 
1/10 - 1/100 of those obtained above, so we here take the capture rate of
dark matter particles in the range of 10-100 GeV 
to be in the range of $ \sim 10^{14} - 10^{16} s^{-1} $ in case of a 
$ 33 \% $ Fe-FeS core, and from $ \sim 10^{12} - 10^{14} $ in case of a $
2 \% $ Fe-FeS core for particles in the range of 10-100 GeV..
During clump core passage, lasting of the order of a year,
the capture rate will rise 
by a maximum factor of $ 10^{8} $,
ie $\sim 10^{22} - 10^{24} s^{-1} $. 
In a single year the number of
particles captured is thus  $\sim 10^{29} - 10^{31} y^{-1} $ (33 \% Fe-FeS
core). Thus the critical number ( of order $ 10^{33} $ )
would be crossed in $ 10^{2-4} $ core passages. Given a rate of one
passage every 30 my, this translates into a time scale of 3 billion to 30
billion years. For the case of a 2 \% Fe-FeS core, the time required to
cross the critical number will be longer than the lifetime of the Solar
System ( $ \sim 4 \times 10^9 yr $ ) since the capture rate is only 
$\sim 10^{27} - 10^{29} y^{-1} $ (2 \% Fe core). Thus, in case of a
heavier core with greater iron content the critical number of dark matter
particles can be crossed within the liftime of the Solar System. This is
the capture rate for iron. That due to other elements will be much less,
however, being off-resonance, and we take the approximation that the
entire capture rate as being that of iron. 
   
\section{Thermal Output}

Depending on the nature of the dark matter ( neutralino, 
photino, gravitino, sneutrino, Majorana neutrino, etc. ), 
different annihilation channels are possible 
\cite{gaisser, bengston}. 
Generally the most significant channels are
$ \chi \bar{\chi} \rightarrow  q \bar{q} $ 
( quark-antiquark ), 
$ \chi \bar{\chi} \rightarrow \gamma \gamma $ 
( photons ) and
$ \chi \bar{\chi} \rightarrow l \bar{l} $ 
( lepton-antilepton ).  
In the quark channel hadronization will take place through 
jets and subsequent radiative decay will lead to mesons 
which in turn will decay through their available channels. 
Hence \cite{bengston} :
\begin{equation}
\chi \bar{\chi} \rightarrow q \bar{q} \rightarrow 
  ( \pi^{0}, \eta, ... ) 
  \rightarrow \gamma + Y  
\end{equation} 

\noindent All annihilation processes 
which directly or indirectly create
photons and where energy is delivered to the core through 
inelastic collisions would lead to the generation of heat 
in the core. Note that the study of the neutrinos is the 
basis for several ongoing dark matter detection experiments 
mentioned above. Here we wish to study this heat generation 
in Ganymede's core through annihilation. This heat is :
\begin{equation} 
\dot{ Q_{G} } = e \dot{ N_{G} } m_X 
\end{equation} 

\noindent where 
$ e $ is the fraction of annihilations which lead to the 
generation of heat in the core of Ganymede.
Note that $ e $ may be as large as unity for the ideal case where
the WIMPs annihilate predominantly through photons only. We
shall however take it to be $ \sim 0.5 $ \cite{gaisser, bengston}
for an order of magnitude estimate. 

Using the capture rates obtained above, the heat output 
due to resonant capture by iron  

\begin{equation}
Q_{ G } = 3.63 \times 10^{8} e\,f\,W 
\end{equation}

Taking specific model results of $ e = 0.5 $, and 
$ f = 0.02 - 0.33 $ for an $ Fe-FeS $ core and  
$ f =  0.014 - 0.26 $  for an $ Fe $ core, the heat output becomes

\begin{eqnarray}
Q_{ G }^{Fe}  = & 3.63 \times 10^{6} 
                  - 6.0 \times 10^{7} W & {\it Fe-FeS\, core} \nonumber \\
   = & 2.54 \times 10^{6} - 4.7 \times 10^{7} W & {\it Fe\, core}
\end{eqnarray}


Thus the heat output is in the range of $ 10^{6-7} W $ for a uniform dark
matter distribution and on iron resonance.  For masses other than 52 GeV
which are non-resonant, the heat output is still 1/10 - 1/100 of that due
to iron.  

\subsection{ Current Heat Output }

The current heat output of Ganymede was measured to be $ \sim 10
mW/m^2 $ at the rock - ice interface (corresponding to a radius of 1800
km) during the Galileo mission
\cite{schu}, i.e. $\sim 4 \times 10^{11} W $.  Thus the heat production
even due to a uniform background of dark matter yields a non-negligible
fraction, in the range $ 0.001 - 0.01 \% $ of current Ganymedean heat
output.  This heat output of $ 10^{6 - 7} W $ rises to $ 10^{14 - 15} W $
in case of clumping by a $ 10^{8} $ factor. 
It is to be noted that the product of period and clumping factor must be
unity, ie. $ \rho_{x} \times \tau = 1 $.
The current heat output of
Ganymede is $ 4 \times 10^{11} $ W. So, if Ganymede is passing through a
dark matter concentration of $ 10^5 $, the current level of Ganymedean
heat output can be explained.  Conversely we can say that the
current level of clumping is less than $ 10^5 $. 


Clump passage lasts for 
a time of the order of 1 year, hence the heat produced due to 
the passage of a clump core in 1 year is 
$\sim 10^{6-7} W \times 10^7 s/year \sim 10^{13-14} J $.
Thus, once every $ \sim 100 my $, Ganymede would experience a jolt of 
$ 10^{13-14} W $, 100 - 1000 times larger than its current heat output.

\subsection{ Formation and Segregation }

The latent heat of fusion of the Earth's mantle is 
$ L^{\oplus}_{mantle} = 420 kJ/kg $ while that of the Earth's
core is $  L^{\oplus}_{core} = 580 kJ/kg $.
Hence a value of $ 500 kJ/kg $ for the latent heat of fusion of 
the Ganymnedean core may not be unreasonable.
For the range of the masses of the Fe and Fe-FeS core used 
above, we find that the latent heat of fusion of the 
Ganymedean core is $ \sim 10^{27} -  10^{28} J $
The specific heat of the Earth's core is 
$ c^{\oplus core}_{p} = 500 - 700 J/kg/K $, while that 
for alpha iron at $ 25 K $ is $ 451J/kg/K $\cite{crc}.
Hence it would not be too unreasonable for the Ganymedean 
core to have a specific heat of $\sim 500 J/kg/K $.
Thus the heat required to raise the temperature of the Ganymedean 
core by one degree is
$ \sim 10^{24} - 10^{25} J $. Thus the heat required for 
core seregation is 
$ \sim 10^{27} - 10^{28} J $.

We already found that the heat generation per year due to a dark 
matter clump core passage through Ganymede yields
$ \sim 10^{13} - 10^{14} J $.
This is hence not sufficient to explain the formation of the Ganymedean
core. Indeed, the results from the Galileo mission indicated that
accretional and radiogenic heating would be sufficient to explain the
formation of Ganymede. The greater problem is that there is no heat source
for the magnetic field.

 The heat evolved is not very large, but still non-negligible. Thus, at
the peak of the iron resonance, the total mass of WIMPs captured per iron
nucleus during the lifetime of the solar system is :

\begin{equation}
0.72 \xi ( \infty ) \sqrt{ \frac{6}{\pi} } \sigma \bar{\rho} \bar{v}
     \frac{v_{esc}^2}{ \bar{v}^2 } \left< \hat{\phi} \right>
\end{equation}

 where $ \bar{\rho} $ is the mean ambient wimp density in the universe, $
v_{esc} $ is the escape velocity of Ganymede and $ \bar{v} $ is the
velocity dispersion of the dark matter particles. Thus, if all the wimp
mass is converted to heat energy, the thermal energy per iron nuclues is

\begin{equation}
 E = \sigma \bar{\rho} \bar{v} \frac{ v_{esc}^2 }{ \bar{v}^2 }
     \left< \hat{ \phi } \right>
\end{equation}

whcih yields $ 1.4 \times 10^{-4} eV $. This may appear small, but it is
in fact equivalent to the heat produced above. In fact, only 6 eV would be
sufficient to explain the melting of the core. However, the process of 
formation and segregation is not explainable by dark matter (the other
sources gravitational and radioactive, were much more important), but
perhaps the magnetic field and a fraction of present heat output
(considered anomalously high) may be due to dark matter annihilations.

 It may appear that the quantity of heat produced in the Earth would also
be very small. However, in this case the larger mass of the body is much
larger, leading to greater quantities of heat that are sufficient to lead
to volcanism. Such drastic effects would not occur in case of Ganymede,
but dark matter could still re-invigourate the magnetic field. 

\subsection{ Ganymedean Magnetic Field }
 
During the last clump core passage
the planet is likely to have received a 
jolt of heat which would have led to re-melting of part of
the core. Convection may have re-started if it had stopped since 
the previous core passage. Planetary magnetic fields are thought
to result from vigorous convection in the interior. Thus this jolt
may have led to the re-starting/envigoration of the planet's
magnetic properties. If the heat could escape quickly, 
then the field may die down before the next clump core passage 
after $\sim O(10 My)$. If not, then the magnetic field would 
continue throughout, becoming
stronger during core passages.  Sustained core convection
and maintenance of the liquid state may have been powered by dark 
matter.


 Gravitational and radiogenic heating are significant only
during the intial phases of a planet's history, gradually declining in
importance as the remnant accretional heat dissipates or the highly active
radionuclides decay. In case of small planets, and especially in case of
Ganymede, the gravitational heat dissipates away much faster than for
larger planets. The only other source of heat that can be significant in
later phases of a planet's history so far has been tidal heating. Thus,
whenever experimental data indicate excess heat production during the
later phases of a planet's history, tidal heating is the refuge. Even when
the present orbital parameters preclude any tidal heating in the known
past, sometimes artificial models are set forth.

\section{ Comparison with Other Heat Sources }
 
Whereas the other heat sources like gravitational heating and radioactive
heating were active during the early part of Ganymede's history and were
much stronger then, dark matter heating is larger now. Moreover, dark
matter provides periodic pulses of heat that cannot be provided by
other heat sources. Tidal heating, although large in the case of Io, is
thought to be small in the case of Ganymede. Models involving jolts of
tidal heating in the past are speculative.
Thus, although dark matter
heating is small, yet it may explain the survival of the magnetic field.
Thus heat through dark matter annihilations appears to be a viable
explanation for the survival of the Ganymedean magnetic field.



\section{ Earth Volcanism}

 It has been shown that the heat produced in Ganymede is not sufficient
to produce the core and represents only less than 1 \% of current heat
output. The question naturally arises as to whether this invalidates the
idea of Earth volcanism set forth in \cite{vdm} ? The answer is that Earth
volcanism is still viable for the following reasons:

\begin{itemize}
 \item The capture rate for large bodies is much larger than just a simple
scaling of masses would suggest. Thus, although the capture rate is linear
in the mass of the body, the term consisting of the square of the escape
velocity is dominant and leads to a much decreased capture rate for
smaller bodies.
 \item Iron represents one-third of the Earth's mass, while current
experimental limits for Ganymede's iron fraction range from 2 \% to 33
\%. The heat output is linear in this term.
 \item Ganymede lies in the potential well of Jupiter, so incident WIMPs
move faster and hence become more difficult to capture. This will decrease
the capture somewhat in case of Ganymede. For the Earth, however, Gould
showed that this decrease in capture is almost fully compenstaed by
enhanced capture of WIMPs bound in solar orbits  \cite{grav}.
\end{itemize}

The passage of a clump core through the Solar System 
would be expected to have consequences not limited to Ganymede 
alone. It has been demonstrated by 
S.Abbas \& A.Abbas \cite{vdm} that the passage of 
the Earth through dark matter clumps leads to the 
heating of the core and the core-mantle boundary ( CMB ). After 
a critical stage, this layer breaks up, ejecting enormous 
superplumes that rise through the mantle. On arrival at the 
surface, these plumes cause
volcanism and attendant mass extinctions recorded 
in the history of the 
Earth. The authors found that the periodicity of clump passages 
approximately agrees with 
that of major flood basalt volcanic episodes as well as 
that observed for mass extinctions. Once again, the model
of dark matter annihilations in planets can exlain another 
puzzle, viz. why the Earth should experience periodic 
episodes of massive volcanism when this planet has in fact 
been cooling ever since its birth.


Hence the heat generation
in the planets and their satellites due to dark matter 
annihilations should be treated as a new source of heat 
in additon to accretional heating, gravitational
heat, radiogenic heating and tidal heating.
The limit of the evaporation mass of the dark matter 
\cite{gould, rich} indicates that in addition to Mercury 
and Ganymede, Venus, Mars, Europa, Callisto, Io and 
Titan should also have
observable heat generation through the above process. One
may even ask whether the dark matter heating mechanism has 
anything to say regarding cryovolcanism \cite{frankel}?
More data would be needed to sort things out. 
This work is in progress 
\cite{prep}.

\section*{ Acknowledgements }

 We thank an anonymous referee for critical and useful comments.


\newpage



\begin{thebibliography}{99}\frenchspacing
\addtolength{\itemsep}{-.05in}
%
\bibitem{silk} J.Silk and A.Stebbins, 
 {\it Astrophys.J.} {\bf 411} (1993) 439 

\bibitem{kolb} E.W.Kolb and I.I.Tkachev,
 {\it Phys.Rev.D } {\bf 50} (1994) 769 

\bibitem{hogan} C.J.Hogan \& M.J.Rees,
 {\it Phys.Letts.B} {\bf 205} (1988) 228

\bibitem{jung} G.Jungmann, M.Kamionkowski \& K.Griest,
 {\it Phys.Rep.} {\bf 267} (1996) 195

\bibitem{press} W.H.Press \& D.N.Spergel
 {\it Astrophys.J.} {\bf 296} ( 1985 ) 679

\bibitem{krauss} L.M.Krauss, M.Srednicki \& F.Wilczek
 {\it Phys.Rev.D } {\bf 33} ( 1986 ) 2079

\bibitem{gould} A.Gould, 
 {\it Astrophys.J.} {\bf 321} (1987) 571 

\bibitem{freese} K.Freese ,
 {\it Phys.Lett.B } {\bf 167} (1986) 295

\bibitem{collar} J.I.Collar,
 {\it Phys.Lett.B } {\bf 368} (1996) 266

\bibitem{gaisser} T.K.Gaisser, G.Steigman \& S.Tilav,
 {\it Phys.Rev.D } {\bf 34} (1986) 2206

\bibitem{bengston} H.-U.Bengtsson, P.Salati \& J.Silk
  {\it Nucl.Phys.B} {\bf 346} ( 1990 ) 129

\bibitem{bere} V.S.Berezinsky,
 {\it Nucl.Phys.B Proc.Suppl.} {\bf 31} (1993) 413 

\bibitem{rich} J.Rich 
 {\it Astropart.Phys.} {\bf 4} (1996) 387 

\bibitem{kiv} M.G.Kivelson, K.K.Khurana, C.T.Russell, R.J.Walker,
  J.Warnecke, F.V.Coroniti, C.Polanskey, D.J.Southwood \& G.Schubert,
  {\it Nature } {\bf 384} (1996) 537

\bibitem{and} J.D.Anderson, E.L.Lau, W.L.Sjogren, 
  G.Schubert \& W.B.Moore,
  {\it Nature} {\bf 384} (1996) 541  

\bibitem{schu} G.Schubert, K.Zhang, M.G.Kivelson, \&
  J.D.Anderson, 
  {\it Nature} {\bf 384} (1996) 544  

\bibitem{kerr} R.A.Kerr, {\it Science} {\bf 273} (1996) 311

\bibitem{jeanloz} R.Jeanloz, D.L.Mitchell, 
  A.L.Sprague, I.de Pater,
  {\it Science} {\bf 268} (1995) 1455

\bibitem{rob} M.S.Robinson \& P.G.Lucey,
  {\it Science} {\bf 275} (1997) 197

\bibitem{crc} CRC Handbook of Chemistry and Physics 1987-1988
  ed. R.C.Weast p.D-178

\bibitem{vdm} S.Abbas \& A.Abbas,
  {\sl ` Volcanogenic Dark Matter and Mass Extinctions '},
  {\it Astropart.Phys.} {\bf 8} (1998) 317-320

\bibitem{grav} A.Gould,
  {\it Astrophys.J.} {\bf 368} (1991) 610-615

\bibitem{frankel} C.Frankel,
  {\sl ` Volcanoes of the Solar System '} 
  Cambridge University Press, Cambridge, U.K. 1996; \\  
  P.Cattermole,
  {\sl ` Planetary Volcanism '} J.Wiley \& Sons,
  Chichester, U.K. 1996

\bibitem{prep} S.Abbas, A.Abbas and S.Mohanty, under preparation  

\end{thebibliography}
\end{document}